# Search for stable and low-energy Ce-Co-Cu ternary compounds using machine learning

Weiyi Xia[1,2], Wei-Shen Tee[2,1], Paul Canfield[1,2], Rebecca Flint[1,2], and Cai-Zhuang Wang[1,2,*]

[1]*Ames National Laboratory, U.S. Department of Energy, Iowa State University, Ames, Iowa 50011, USA*

[2]*Department of Physics and Astronomy, Iowa State University, Ames, Iowa 50011, USA*

* wangcz@ameslab.gov

**Abstract**

Cerium-based intermetallics have garnered significant research attention as potential new permanent magnets. In this study, we explore the compositional and structural landscape of Ce-Co-Cu ternary compounds using a machine learning (ML)-guided framework integrated with first-principles calculations. We employ a crystal graph convolutional neural network (CGCNN), which enables efficient screening for promising candidates, significantly accelerating the materials discovery process. With this approach, we predict five stable compounds, $Ce_3Co_3Cu$, $CeCoCu_2$, $Ce_{12}Co_7Cu$, $Ce_{11}Co_9Cu$ and $Ce_{10}Co_{11}Cu_4$, with formation energies below the convex hull, along with hundreds of low-energy (possibly metastable) Ce-Co-Cu ternary compounds. First-principles calculations reveal that several structures are both energetically and dynamically stable. Notably, two Co-rich low-energy compounds, $Ce_4Co_{33}Cu$ and $Ce_4Co_{31}Cu_3$, are predicted to have high magnetizations.

## 1. Introduction

Ce-based intermetallic compounds are of great interest due to the relative abundance of Ce compared to other rare earths, and the potential technological applications of Ce-based materials in clean energy. Currently, the most widely used high performance magnets for energy generation/conversion and information storage, like generators, motors, mobile machines and computer hard drives, contain critical rare earth (RE) elements like Nd, Sm, and Dy [1-2]. Owing to insecure sources and supply of these critical RE materials, there are considerable interests and efforts in replacing these critical RE elements with more abundant and cheaper Ce to meet the performance and cost goals for advanced electro-magnetic devices [3-7]. Ce-based intermetallic compounds are also interesting from the fundamental science point of view due, as they can vary in valence from the non-magnetic $Ce^{4+}$ to magnetic $Ce^{3+}$, and form heavy fermion and mixed valent materials with exotic quantum criticality, superconductivity and magnetism [19-20]. Because of the technological and fundamental importance of Ce-based intermetallic compounds, comprehensive knowledge about the stable and low-energy, potentially metastable, phases of these compounds and alloys is highly desirable.

In this paper, we focus on searching for the stable and low-energy (100 mev above the convex hull and therefore potentially metastable) phases of Ce-Co-Cu ternary compounds as a step toward this goal. This ternary system is chosen motivated by experimental studies which suggest ternary Ce-Co-Cu intermetallic compound/alloys would be a promising class of materials for discovery and design of novel strong permanent magnet with light RE elements. For example, $SmCo_5$ has been known to be a strong permanent magnet since the early 1960s. However, Samarium–cobalt magnets are expensive and subject to supply and price fluctuations. It has been shown that replacing Sm with much cheaper and abundant Ce and partially substituting Co with non-magnetic element Cu can result in Ce-Co-Cu compounds, such as $CeCo_4Cu$ alloy, which exhibit desirable magnetic properties for permanent magnet applications [8-18]. A recent combination experimental and first-principles calculation study on Cu dopped $SmCo_5$ [21] show that Cu doping can enhance the stability and magnetic anisotropy of the compound. The study suggests that Cu doping modifies the crystal field splitting (CFS) of the RE atoms thus the magnetic anisotropy. Cu doping also modify the exchange coupling between the RE atoms and Co atoms in favor of stronger magnetic anisotropy. Moreover, $Ce_2Co_{17}$ has high magnetization, but the magnetic anisotropy is too low for permanent magnet application. However, it has also been shown by experiment that by replacing some Co with Cu atoms to form $Ce_2Co_{17-x}Cu_x$ (x=1) can substantially increase the magnetic anisotropy [22]. Therefore, search for more stable and low-energy Ce-Co-Cu compounds and alloys it is interesting and highly desired.

To our best knowledge, there are no known stable Ce-Co-Cu ternaries, even though there are a number of interesting experimentally known compounds with some amount of Co-Cu site disorder such as $CeCo_{5-x}Cu_x$, $Ce_2Co_{16}Cu$, [8-18, 23]. We want to search to see if we can find new stable materials, or low-energy materials that might be similarly stabilized by site disorder. We are particularly interested in materials with high Co concentrations, as these are particularly promising for developing new permanent magnets. Owing to a considerable large number of possible combinations of composition ratio among the three elements and the potential crystal structures they may take, we anticipate more stable and low-energy Ce-Co-Cu ternary phases remain to be discovered. Nevertheless, the vast combinatorial composition-structure space presents challenges for the discovery of the new phases. To overcome this difficulty, recent machine learning (ML) approaches have been employed [24-36], which can greatly accelerate materials design and discovery. In this work, we utilize a machine learning (ML) guided framework [38] to efficiently select promising candidates for first-principles calculations. Such a ML-guided *ab initio* approach greatly accelerates the materials discovery process and enables us to predict 5 new Ce-Co-Cu ternary compounds with their formation energies below the currently known convex hull. We also predict 250 low-energy low-energy ternary compounds for Ce-Co-Cu system. We hope these ML-guided computational predictions can motivate experimental efforts in making them. If any of these structures can be experimentally synthesized, they would be the first stable ternary Ce-Co-Cu compounds without site disorder. In many cases, occupation disorder among the Co and Cu atoms may further stabilize the structure, which would require further investigations.

The rest of the paper is organized as follows. In section 2, we describe how ML is used to accelerate the discovery of new compounds. The first-principles calculation results guided by ML are presented in Section 3. In section 4, we assess the dynamical stability of the predicted stable and low-energy interesting low-energy compounds. Finally, a summary is given in section 5.

## 2. Efficient structure searches guided by ML-guided framework

A schematic sketch of the ML-guided framework used in this study to search for low-energy stable and low-energy Ce-Co-Cu ternary compounds is shown in **Fig. 1**. A crystal graph convolutional neural network (CGCNN) ML method [24] is used in this framework. In CGCNN, a crystal structure is represented by a crystal graph which encodes both the atomic information of the atoms and the bonding interactions among them in the crystal. A convolutional neural network is used to process the crystal graphs to optimize the descriptors followed by another neural network to map out the relationship between the descriptors and the physical properties of the crystals. The training data in CGCNN can be generated by first-principles calculations, which enables a sufficient volume of data for the supervision training. In this study, we use a CGCNN ML model to predict the formation energy ($E_f$) for Ce-Co-Cu ternary compounds. The CGCNN model for the formation energy predictions of compounds developed by *Xie* and *Grossman* [24] was adopted as the 1$^{st}$ generation CGCNN (1G-CGCNN) in our framework. This model was trained using the first-principles calculation results of structures and energies of 28,046 binary and ternary compounds from the Materials Project (MP) database [23]. These 28,046 binary and ternary compounds contain a wide range of chemical elements in the periodic table. Therefore, the 1G-CGCNN model is not material-specific and can be used as a good starting point for any compound. After the structure candidates selected by 1G-CGCNN model are optimized by first-principles calculations, we obtained 1716 Ce-Co-Cu ternary compounds. Then we use these 1716 structures to train a 2nd generation CGCNN model (2G-CGCNN) specifically for predicting the formation energy ($E_f$) of Ce-Co-Cu compounds. The 2G-CGCNN is then applied to search for more promising ternary Ce-Co-Cu compounds. It should be noted that such a CGCNN approach is limited to known structure types in the database. Stable structures whose structure motifs are not present in the existing structural databases will be missed. If a new compound/composition is found experimentally to be stable but missed by the current algorithm (say a new structure type), then this result changes the hull and predictions.

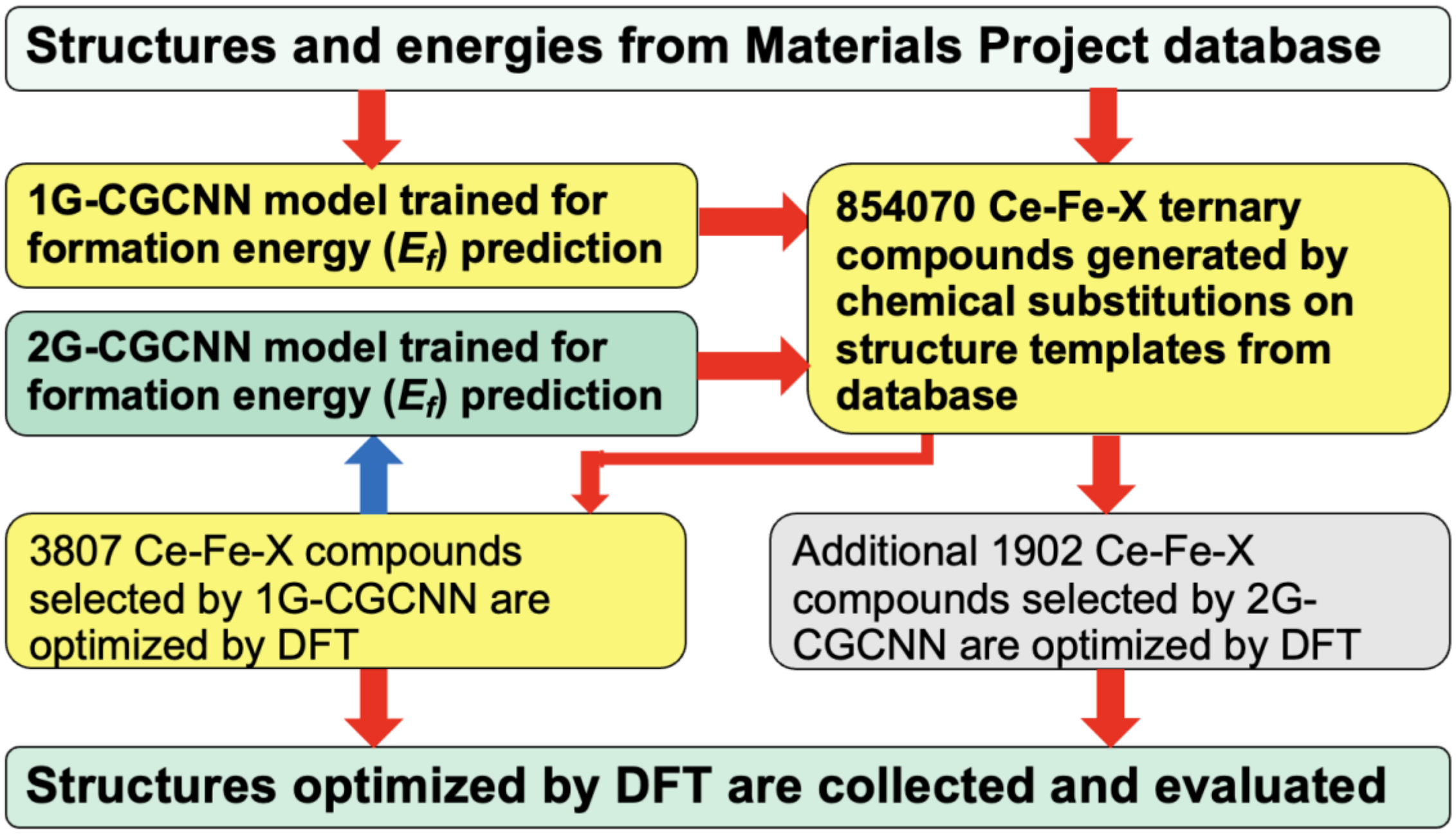

**Fig. 1.** A schematic flowchart of ML-guided framework for efficient discovery of stable and low-energy Ce-Co-Cu ternary compounds. In addition to training the 1G-CGCNN, a structure pool of hypothetical ternary Ce-Co-Cu compounds are generated by substitution of Ce, Co and Cu on the atomic-sites of 28469 ternary compounds extracted from the MP database [23]. For a given ternary template, there are six ways to shuffle the three elements Ce, Co, and Cu on the atomic positions of the structure. We also allow the volume of the unit cell to vary by a scaling factor of 0.96 to 1.04, in increments of 0.02, to help the CGCNN model differentiate the energetic stability of the same structure with different bond lengths. In this way, a pool of 854,070 hypothetical ternary Ce-Co-Cu compounds is generated as shown in **Fig. 1**.

The 1G- and 2G-CGCNN energy models are applied respectively to the same structure pool of 854,070 ternary Ce-Co-Cu compounds. The formation energy distribution (histogram) from the predictions of the 2 generation CGCNN models are shown in **Fig. 2.** The formation energy $E_f$ per atom is defined relative to the elemental phases of a $Ce_\alpha Co_\beta Cu_\gamma$ with $\alpha+\beta+\gamma=1$ as

$$E_f = E(Ce_\alpha Co_\beta Cu_\gamma) - \alpha E(Ce) - \beta E(Co) - \gamma E(Cu).$$

Here, $E(Ce_\alpha Co_\beta Cu_\gamma)$ is the total energy per atom of a $Ce_\alpha Co_\beta Cu_\gamma$ structure. Reference energies are the total energies per atom of face-centered cubic Ce, hexagonal close-packed Co, and face-centered cubic Cu. Based on the $E_f$ histograms shown in **Fig. 2** and after removing the redundancy structures with similarity we select 3807, and 1912 structures, respectively, from 1G- and 2G-CGCNN predictions for further evaluation by first-principles calculations.

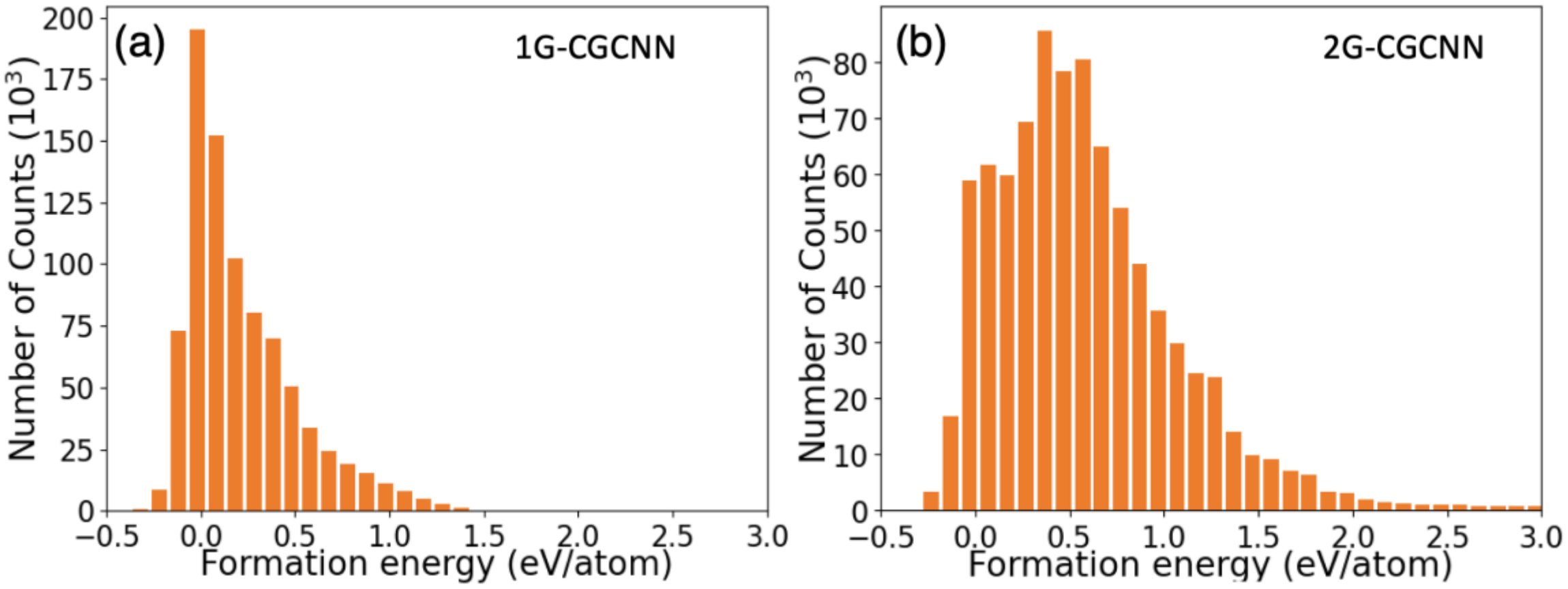

**Fig. 2.** Distribution of formation energies ($E_f$) of the hypothetical ternary Ce-Co-Cu ternary compounds predicted from the (a) 1G, and (b) 2G CGCNN energy models. The total number of structures is 854,070. We select 3807 and 1912 structures from 1G- and 2G-CGCNN predictions respectively for further optimization by DFT calculations.

## 3. First-principles calculation results

The first-principles calculations are performed based on density functional theory (DFT) using the VASP package [37-38], with Perdew-Burke-Ernzerhof (PBE) functionals [39] combined with the projector-augmented wave (PAW) method [40] and a cutoff energy of 520 eV. We use a $k$-point grid with a mesh size of $2\pi \times 0.025$ Å$^{-1}$ generated by the Monkhorst-Pack scheme. This mesh size is fine enough to sample the first Brillouin zone for achieving better $k$-point convergence [41]. The lattice vectors and the atomic positions of candidate structures selected from 1G- and 2G-CGCNN predictions are fully optimized by the DFT calculations until forces on each atom are less than 0.01 eV/atom. In these DFT calculations, spin-orbit interactions are neglected, which are not expected to have a significant effect on the formation energies.

The results from the DFT calculations show that 1716 and 662 non-equivalent structures selected from the 1G- and 2G-CGCNN predictions can be fully optimized. Other structures that cannot pass the electronic self-consistent calculations or are duplicate structures are discarded. These discarded structures are most likely to be far from the realistic structures for Ce-Co-Cu ternary compounds. There are 2378 structures in total identified as new structures for ternary Ce-Co-Cu compounds. The 1716 structures from 1G-CGCNN are used to train the 2G-CGCNN model specifically for Ce-Co-Cu ternaries, as discussed above. To evaluate the thermodynamic stability of these newly predicted ternary compounds, we then calculate the formation energies ($E_{hull}$) of these structures with respect to the Ce-Co-Cu ternary convex hull at the accuracy level DFT. The $E_{hull}$ is the decomposition energy of a $Ce_{\alpha}Co_{\beta}Cu_{\gamma}$ ternary compound with respect to the nearby three known stable phases, which can be ternary, binary or elemental phases. The chemical compositions of these phases are located at the vertexes of the Gibbs triangle that encloses the composition of the $Ce_{\alpha}Co_{\beta}Cu_{\gamma}$.The compositions of stable and low-energy low-energy ($E_{hull} \leq 0.1$ eV/atom) Ce-Co-Cu ternary phases with respect to the currently known convex hull predicted from our CGCNN+DFT approach are shown in **Fig. 3 (a)** and **(b)** respectively. More detailed information about these structures is shown in Table S1 in the Supplementary Materials.

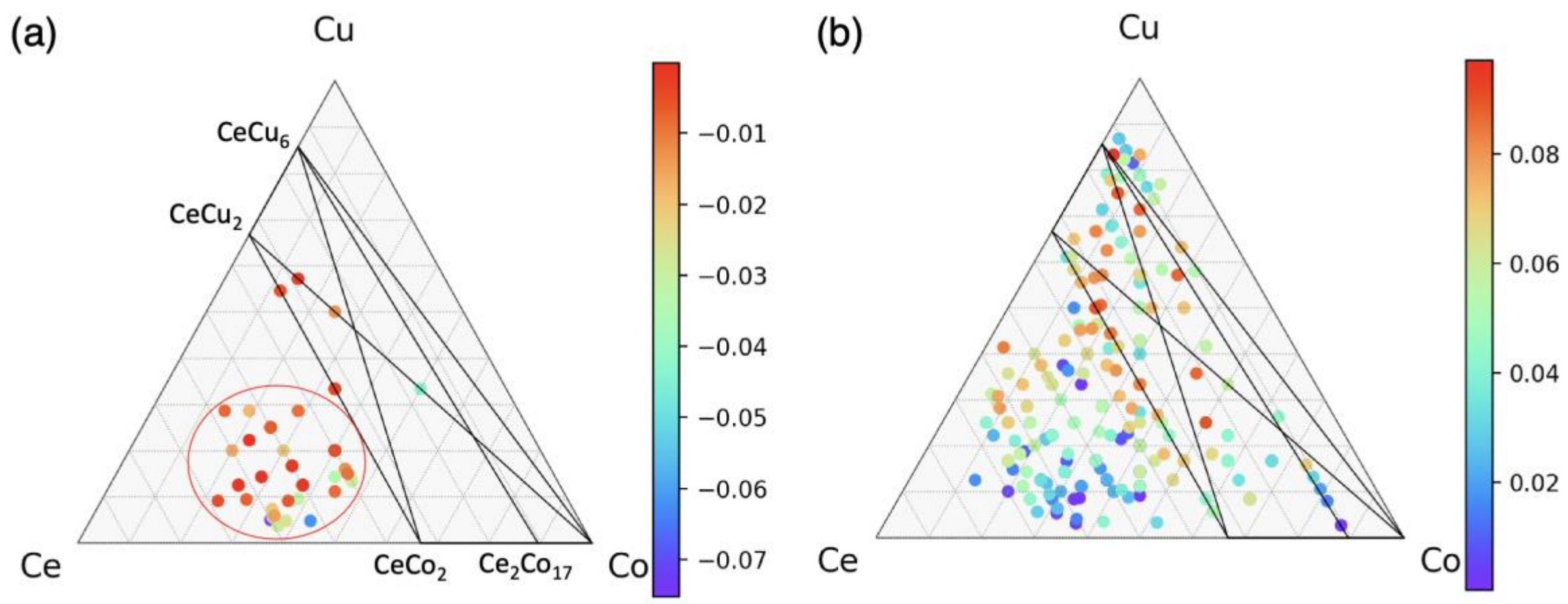

**Fig. 3.** The compositions of (a) stable and (b) low-energy ($E_{hull} \leq 0.1$ eV/atom) Ce-Co-Cu ternary phases with respect to the currently known convex hull predicted from our CGCNN+DFT approach. The compositions in the convex are colored by the lowest-$E_{hull}$ for this composition. The $E_{hull}$ shown in the color bars are in the unit of eV/atom. The predicted stable structures are enclosed by a red circle as shown in bottom left region in (a).

From the results shown in **Fig 3** and **Table S1**, we can see that many structures are predicted to be stable or good low-energy (with formation energies under the convex hull or within 100 meV/atom above the convex hull) based on currently known stable binary and elementary phases. We note that the stability presented here is based on DFT energies at T= 0 K, and no partial site occupancies are considered. Allowing partial site occupancies may further lower the free energies at finite temperatures due to the entropy contribution. Additionally, many stable compounds from our predictions cluster in the Ce-rich region of the ternary convex hull, as indicated by the red circle in **Fig. 3(a)**. This suggests that these predicted stable phases may compete during phase selection and stability in synthesis. When we include the energies of all newly predicted structures to construct a new convex hull for the Ce-Co-Cu ternary system, only five phases remain stable: $Ce_3Co_3Cu$, $CeCoCu_2$, $Ce_{12}Co_7Cu$, $Ce_{11}Co_9Cu$ and $Ce_{10}Co_{11}Cu_4$ , as shown in the new convex hull plot in **Fig. 4 (a)**. Notably, these five stable phases have formation energies of 42, 35, 75, 62, and 35 meV/atom below the previously known convex hull, respectively. However, the low Co content in these compounds makes them less suitable for applications in magnetic materials. However, these Ce-rich compounds still might be interesting if the Ce is mixed valent.

Accordingly, the low-energy ($E_{hull} \leq 0.1$ eV/atom) low-energy phases are also redefined as shown in **Fig. 4 (b)**. There are two Co-rich phases with $E_{hull}$ within 0.02 eV/atom above the new convex hull as indicated by the red circle in the bottom-left corner of **Fig. 4(b)**. These Co-rich low-energy structures might be interesting for magnetic materials.

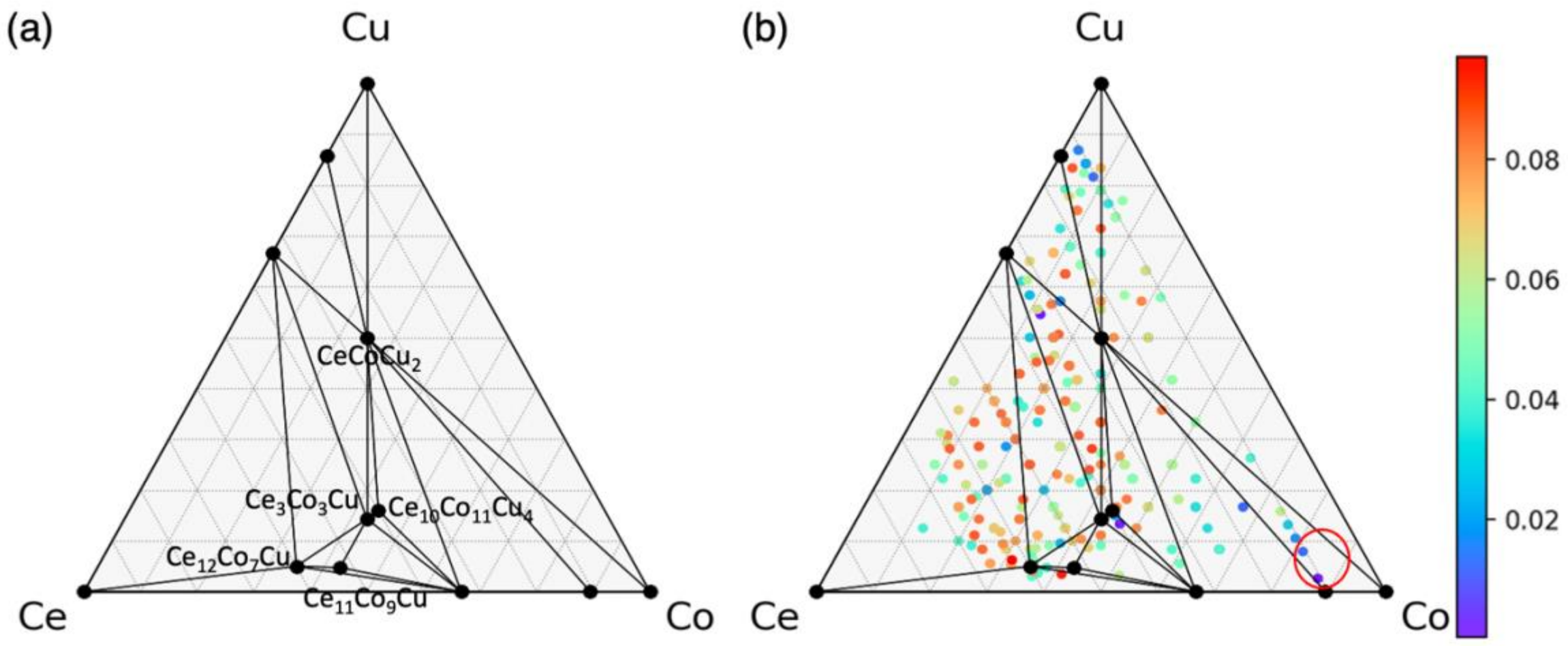

**Fig. 4.** (a) New convex hull for Ce-Co-Cu ternaries after the new structures predicted from the present study are included. (b) Low-energy ($E_{hull} \leq 0.1$ eV/atom) Ce-Co-Cu ternary phases with respect to the new convex hull predicted from our CGCNN+DFT calculations. For a given composition, there may be several stable or low-energy structures. The colors in (b) are shown according to the lowest $E_{hull}$ for the given composition. The $E_{hull}$ shown in the color bars are in the unit of eV/atom. The two low-energy Co-rich compounds (shown later in Fig. 6) are enclosed by a red circle in the bottom-right corner of (b).

The structures of the five stable structures and the two Co-rich low-energy structures obtained from our predictions are plotted in **Fig. 5** and **Fig 6** respectively. Detailed information of these structures is given in **Table 1 below**.

The crystal structure of the $Ce_3Co_3Cu$ compound shown in **Fig 5 (a)** has an orthorhombic lattice with *Cmcm* space group symmetry. There are two Wyckoff positions for the Ce and Co atoms. The first Ce atom is bonded to one Co with bond length 2.23Å and three

Cu atoms with distance of 2.86 Å and 3.18 Å. The second Ce atom is bonded to one Co with a similar 2.24 Å distance and two Cu atoms with 3.43 Å distance. Cu is bonded to eight Ce atoms and four equivalent Co atoms to form a mixture of distorted edge, face, and corner-sharing $CuCe_8Co_4$ cuboctahedra.

**Table 1.** Crystallographic data of $Ce_3Co_3Cu_4$ (space group of *Cmcm*).

| Phase | Lattice param. | Wyckoff site | Atom | x | y | z |
|---|---|---|---|---|---|---|
| $Ce_3Co_3Cu_4$ | *a*=4.367 | 8f | Ce | 0 | 0.2895 | 0.6077 |
| | *b*=8.462 | 4c | Ce | 0 | 0.0049 | 0.25 |
| | *c*=13.728 | 8f | Co | 0 | 0.4483 | 0.0881 |
| | α=β=γ=90° | 4c | Co | 0 | 0.2695 | 0.25 |
| | | 4a | Cu | 0 | 0 | 0 |

$CeCoCu_2$ crystallizes in the orthorhombic *Pnma* space group symmetry, as shown in **Fig 5(b)**. The Ce atom is bonded in a distorted geometry to one Co and 10 Cu atoms with Ce-Co bond length of 2.16 Å and Ce-Cu bond length ranging from 2.92 – 3.12 Å.

**Table 2.** Crystallographic data of $CeCoCu_2$ (space group of *Pnma*).

| Phase | Lattice param. | Wyckoff site | Atom | x | y | z |
|---|---|---|---|---|---|---|
| $CeCoCu_2$ | *a*=7.437 | 4c | Ce | -0.0288 | 0.25 | -0.0853 |
| | *b*=6.950 | 4c | Co | 0.1070 | 0.25 | 0.2992 |
| | *c*=4.958 | 8d | Cu | 0.1658 | 0.5510 | 0.5616 |
| | α=β=γ=90° | | | | | |

**Fig 5 (c)** shows the crystal structure of $Ce_{12}Co_7Cu$. It has a tetragonal lattice with *I4/mcm* space group, with a large unit cell of 80 atoms. There are three inequivalent Ce and Co sites, and a single Cu site. Cu atoms are bonded in a body-centered cubic geometry to eight equivalent Ce atoms, with bond length of 3.10 Å. The bond length between Ce and Co atoms ranges from 2.36 – 2.95 Å.

**Table 3.** Crystallographic data of $Ce_{12}Co_7Cu$ (space group of *I4/mcm*).

| Phase | Lattice param. | Wyckoff site | Atom | x | y | z |
|---|---|---|---|---|---|---|
| | | 32m | Ce | 0.0700 | 0.2071 | 0.1413 |
| | *a*=11.498 | 8g | Ce | 0 | 0.5 | 0.1405 |
| | *b*=11.498 | 8h | Ce | 0.1748 | 0.6748 | 0 |
| $Ce_{12}Co_7Cu$ | *c*=12.815 | 16l | Co | 0.6783 | 0.1783 | 0.6841 |
| | α=β=γ=90° | 8h | Co | 0.6287 | 0.1287 | 0 |
| | | 4a | Co | 0 | 0 | 0.25 |
| | | 4c | Cu | 0 | 0 | 0 |

$Ce_{11}Co_9Cu$ forms in orthorhombic *Iba2* space group with 84 atoms in the unit cell as shown in **Fig. 5 (d)**. There are 6 Wyckoff sites for Ce, 5 for Co and one for Cu. Cu is bonded in a 5-coordinate geometry to 4 Ce atoms and two equivalent Co atoms. The bond lengths are 2.94 and 3.43 Å for Ce-Cu, and 2.47 Å for Co-Cu.

**Table 4.** Crystallographic data of $Ce_{11}Co_9Cu$ (space group of *Iba2*).

| Phase | Lattice param. | Wyckoff site | Atom | x | y | z |
|---|---|---|---|---|---|---|
| $Ce_{11}Co_9Cu$ | *a*=11.012<br>*b*=11.736<br>*c*=12.734<br>α=β=γ=90° | 8c | Ce | 0.5759 | 0.2654 | 0.3532 |
| | | 8c | Ce | 0.5846 | 0.7226 | 0.1384 |
| | | 8c | Ce | 0.1636 | 0.6717 | 0.4939 |
| | | 8c | Ce | 0.6820 | 0.4378 | 0.1319 |
| | | 8c | Ce | 0.2003 | -0.0580 | 0.3616 |
| | | 4b | Ce | 0 | 0.5 | 0.6916 |
| | | 8c | Co | 0.5298 | 0.4009 | 0.4878 |
| | | 8c | Co | 0.1638 | 0.6111 | 0.0070 |
| | | 8c | Co | 0.1718 | 0.3212 | 0.1862 |
| | | 8c | Co | 0.1738 | 0.6760 | 0.3058 |
| | | 4a | Co | 0 | 0 | 0.7570 |
| | | 4b | Cu | 0 | 0.5 | -0.0774 |

Finally, $Ce_{10}Co_{11}Cu_4$ has a monoclinic lattice with *C2/m* space group, as shown in **Fig 5 (e)**. There are 5 inequivalent Ce sites, 6 Co sites and 2 Cu sites. In both Cu sites, Cu is bonded to 8 Ce atoms and 4 Co atoms to form a mixture of distorted face and edge-sharing $CuCe_8Co_4$ cuboctahedra. The bond lengths range from 2.89 – 3.32 Å for Ce-Cu, 2.50 – 2.51 Å for Co-Cu, and 2.24 – 2.29 Å for Ce-Co.

**Table 5.** Crystallographic data of $Ce_{10}Co_{11}Cu_4$ (space group of *C2/m*).

| Phase | Lattice param. | Wyckoff site | Atom | x | y | z |
|---|---|---|---|---|---|---|
| $Ce_{10}Co_{11}Cu_4$ | *a*=10.073<br>*b*=4.321<br>*c*=20.250<br>α=γ=90°<br>β= 98.08° | 4i | Ce | 0.5179 | 0 | 0.6606 |
| | | 4i | Ce | 0.0706 | 0 | 0.1316 |
| | | 4i | Ce | 0.1119 | 0 | 0.4581 |
| | | 4i | Ce | 0.6948 | 0 | -0.0566 |
| | | 4i | Ce | 0.7040 | 0 | 0.2544 |
| | | 4i | Co | 0.0002 | 0 | 0.7620 |
| | | 4i | Co | 0.5942 | 0 | 0.5595 |
| | | 4i | Co | 0.6017 | 0 | 0.0402 |
| | | 4i | Co | 0.1894 | 0 | 0.3578 |
| | | 4i | Co | 0.2168 | 0 | 0.8434 |
| | | 2a | Co | 0 | 0 | 0 |
| | | 4i | Cu | 0.6081 | 0 | 0.8025 |
| | | 4i | Cu | 0.2024 | 0 | 0.6015 |

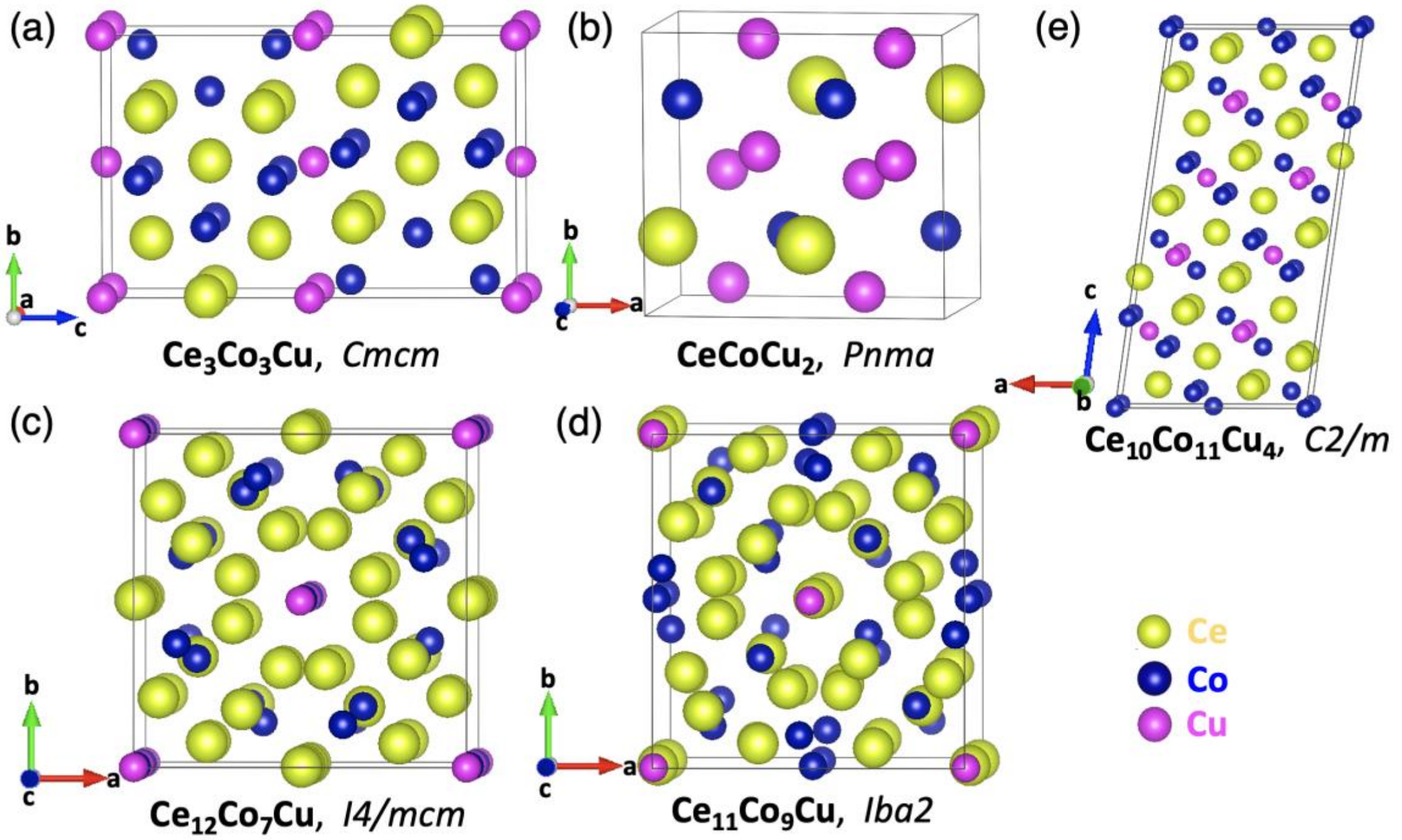

**Fig. 5**. The crystal structures of the 5 stable Ce-Co-Cu ternary compounds from our CGCNN+DFT calculations. The Ce, Co, and Cu atoms are presented by green, dark blue and magenta balls respectively.

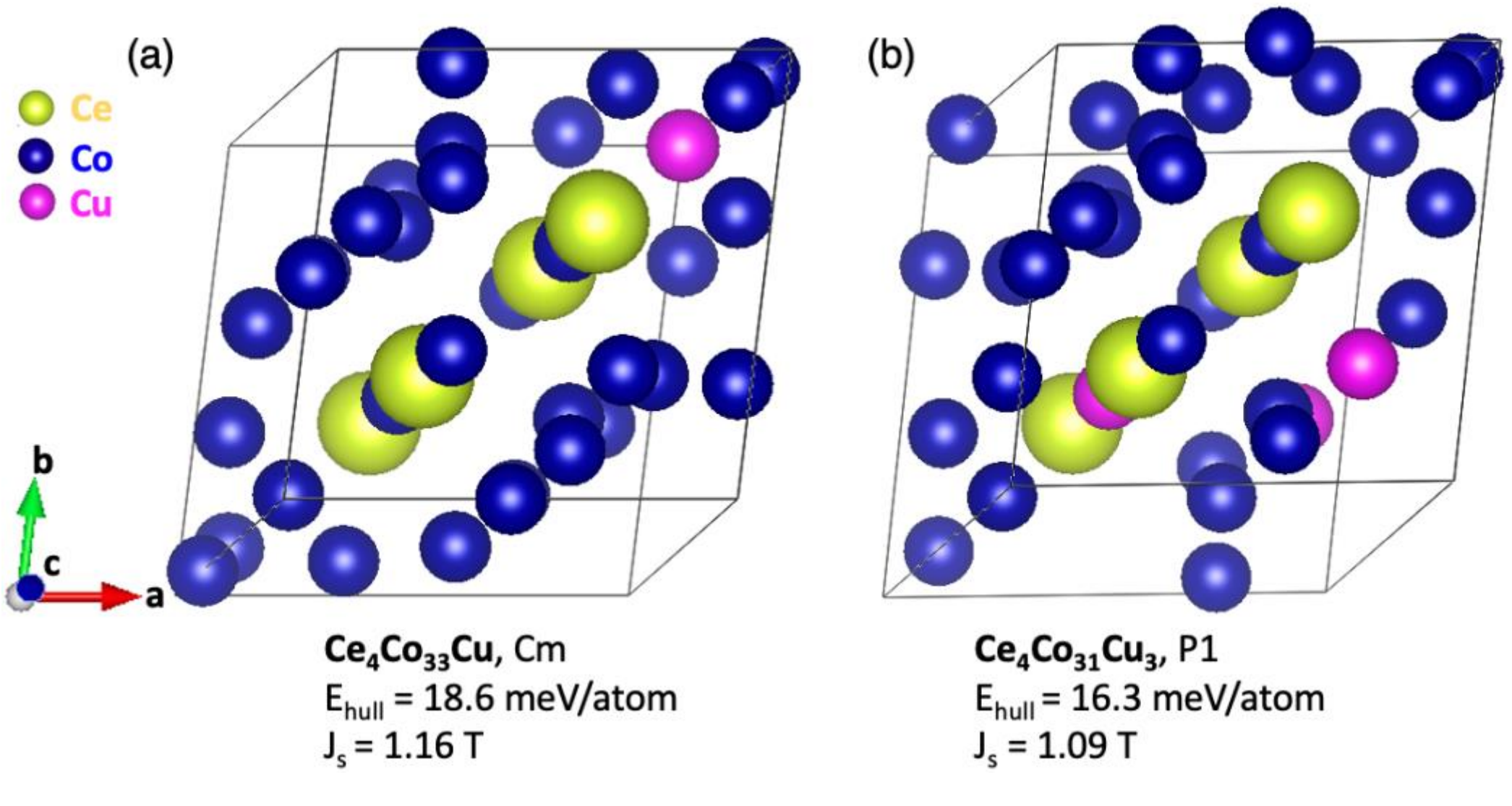

**Fig. 6**. The crystal structures of the two Co-rich low-energy Ce-Co-Cu ternary compounds from our CGCNN+DFT calculations. Both can be viewed as slight distortions of Ce2Co17 with (a) 1 Cu and (b) 3 Cu atoms substituted for Co. The Ce, Co, and Cu atoms are presented by green, dark blue and magenta balls respectively. $J_s$ is the magnetization of the compound in units of tesla (T).

The two Co-rich low-energy compounds $Ce_4Co_{33}Cu$ and $Ce_4Co_{31}Cu_3$ both have monoclinic lattices, with *C2/m* and *P-1* space groups, respectively. Both compounds can be viewed as distorted, Cu-doped variants of the binary $Ce_2Co_{17}$. Despite $Ce_2Co_{17}$'s very high Curie temperature and large saturation moment, it remains a rather inferior magnet due to its small uniaxial magnetic anisotropy energy (MAE). Previous studies have shown that the poor MAE stems from the negative contribution of the Co atoms occupying the "dumbbell" 4f Wyckoff site in the rhombohedral structure. Various attempts have been made to introduce dopants such as Fe, Mn, Al, Zr, etc., into this system to enhance the MAE, aiming to develop potential permanent magnets [42-49]. Similarly, a doped Cu atom in the "dumbbell" site may also help enhance the MAE. Notably, the experimentally synthesized $Ce_2Co_{16}Cu$ has the same hexagonal lattice with the *P6₃/mmc* space group as the binary $Ce_2Co_{17}$, with the Co 4f Wyckoff site half-occupied by Cu [22]. Previous computational work has also evaluated the MAE of $Ce_2Co_{15}Cu_2$, indicating an enhanced magnetocrystalline anisotropy constant $K_1$ of 2 MJ/m$^3$[48]. It will be interesting to investigate the MAE for these two newly predicted structures, where we expect that $Ce_4Co_{31}Cu_3$ will have a higher MAE, as it has one Cu on a dumbbell site Moreover, the higher configurational entropy in the disordered structure with partial dumbbell site occupancy, as in $Ce_2Co_{16}Cu$, can help to stabilize the structures at finite temperature.

We next explore the electronic and magnetic properties of these seven ternary Ce-Co-Cu compounds at the DFT-PBE level. Prior research has shown that Ce is non-magnetic in tetravalent $CeCo_5$ [5, 50], while antiferromagnetic behavior is observed around 4 K in trivalent $CeCu_5$ [6]. Doping Cu into $CeCo_5$ enhances stability, magnetic anisotropy, and coercivity, which can be attributed to the transition from $Ce^{4+}$ to $Ce^{3+}$ [12, 14-15]. Additionally, X-ray absorption spectroscopy suggests a Ce valence between 3.0 and 3.3 in $Ce_2Co_{17}$ [51], indicating weak mixed valency. Our DFT-PBE calculations find non-magnetic Ce and Co behavior in the five stable Ce-Co-Cu compounds, where both exhibit negligible magnetic moments, ranging from (0 to 0.03 $\mu_B$) in both the spin-polarized calculations and multiple antiferromagnetic configurations. Note that Ce magnetization may not be correctly evaluated through DFT due to its limitation in describing Ce 4f electons, as discussed below. In contrast, both Co-rich compounds, $Ce_4Co_{33}Cu$ and $Ce_4Co_{31}Cu_3$ display significant magnetization. Ce possesses a magnetic moment of approximately 0.9 $\mu_B$, oriented opposite to the Co atoms, which have moments ranging from 1.4 to 1.6 $\mu_B$.

The spin-polarized electronic density-of-states (DOSs) for the ferromagnetic state obtained from our DFT-PBE calculations are depicted in **Fig. 7.** The results show that the electronic bands near the Fermi level are predominantly contributed by the d-electrons of Co, with the f-orbitals of Ce located slightly above the Fermi level in all seven compounds. Experimental evidence suggests that $Ce^{4+}$, $Ce^{3+}$ and mixed valence are all possible in this ternary system. Current DFT-PBE calculations do not treat the Ce interactions appropriately and so we cannot draw any conclusions about the valence of the Ce for any of the compounds. The 0th order guess is that the Ce-rich compounds are all non-magnetic (Ce4+ or mixed valent) as none of the magnetic structures indicated Ce moments, by contrast to the Co-rich compounds. Further investigation using advanced methods, such as dynamical mean-field theory (DMFT), is highly recommended.

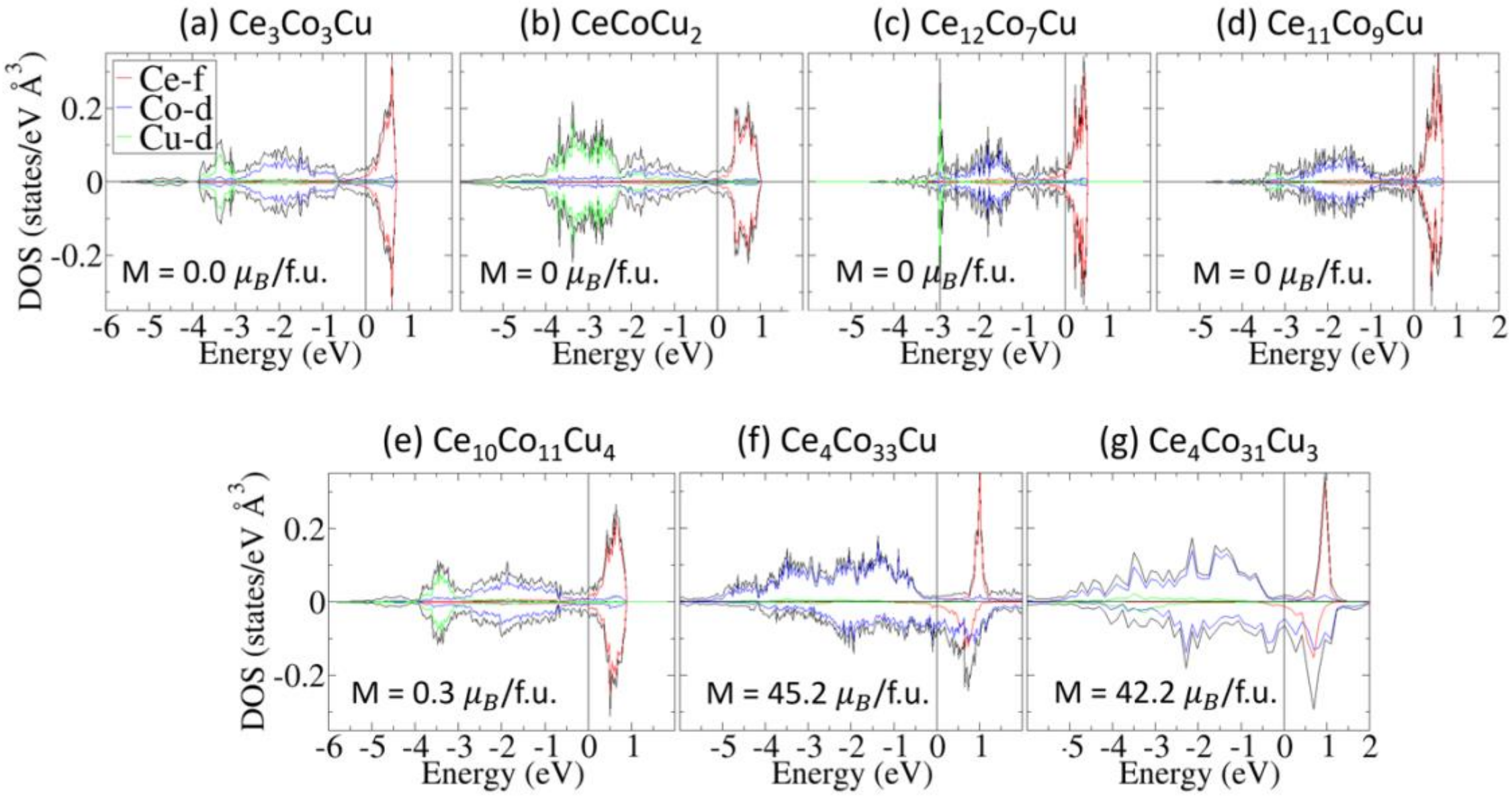

**Fig. 7**. Electronic density of states (DOS) of the 5 stable (a-e) and two low-energy (f-g) Ce-Co-Cu ternary compounds. The Fermi level is shifted to zero. The red line represents the projected DOS of Ce *f* orbitals; the blue line represents the Co *d* orbitals, and the green line represents the Cu *d* orbitals.

## 4. Dynamical stability

To assess the dynamical stabilities of the five predicted stable phases and the two Co-rich low-energy phases, we performed phonon calculations for these seven compounds. The Phonopy package [52-53] was used with the finite displacement method. Supercells are used so that the lattice parameters in each direction are around 14 Å, and the forces were calculated with displacement along different directions. The DFT calculations adopted the same setups used in the structure optimization as well as the formation energy calculations discussed above. Force constants were then calculated from the set of forces. Dynamical matrices were built from the force constants and thus phonon frequencies and eigenvectors were obtained at the specified q points. No imaginary vibrational frequencies were found for three out of five predicted stable structures and the two low-energy structures, indicating that these 5 structures are dynamically stable at T=0 K. The phonon dispersions of these five dynamically stable structures are shown in **Fig. 8**.

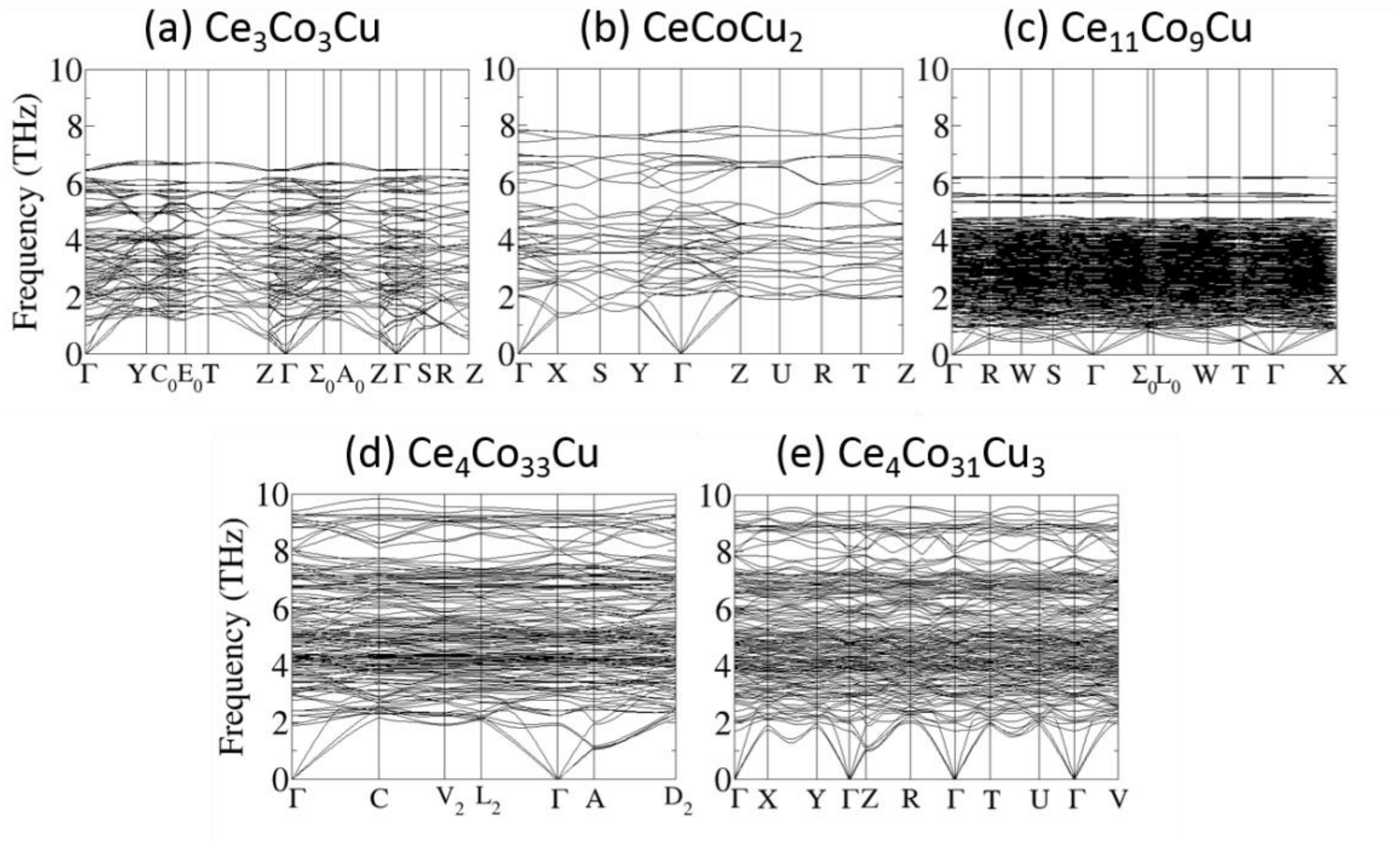

**Fig. 8**. Phonon dispersions of the 5 stable and low-energy Ce-Co-Cu ternary compounds that are dynamically stable.

The $Ce_{10}Co_{11}Cu_4$ and $Ce_{12}Co_7Cu$ structures did have imaginary-frequencies in the harmonic phonon calculation at T = 0 K. To see if these two structures can be stabilized with anharmonic interactions included at finite temperatures, we performed *ab initio* molecular dynamics (AIMD) at a constant temperature of 500 K and constant zero pressure using NPT (constant number of particles, pressure and temperature) ensemble to study their stabilities. A 1x4x1 supercell (200 atoms) for $Ce_{10}Co_{11}Cu_4$ and a 1x1x1 supercell (80 atoms) for $Ce_{12}Co_7Cu$ were used. Only the Γ point was used for sampling the Brillouin zone, with a smaller plane-wave cutoff energy of 300 eV. Both structures were calculated over at least 60 ps. The total energy, pressure, and volume as a function of time at 500K are shown in **Fig. 9**. The total energy, pressure, and the volume of the two crystals fluctuated around their average values for more than 60 ps. These AIMD simulation results indicate that the structures do not collapse or transform into other structures. We also take the atomic position average of the last 10000 ionic steps (30 ps) and find that the average structures are the same as the original structures. These results suggests that these two structures can maintain stability at 500 K.

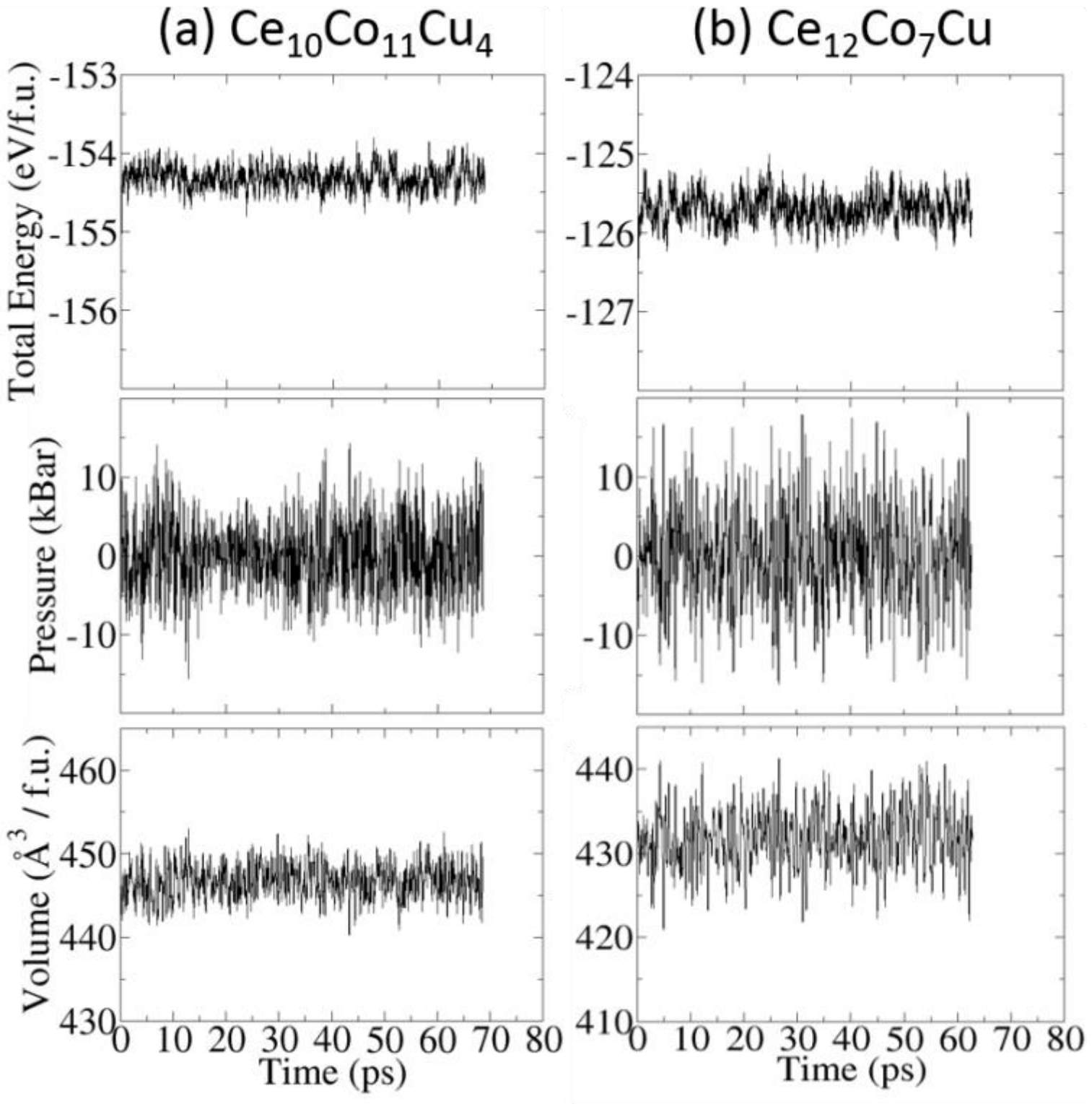

**Fig. 9.** The total energy, pressure, and volume as a function of time at 500K for (a) $Ce_{10}Co_{11}Cu_4$ and (b) $Ce_{12}Co_7Cu$ obtained from ab initio molecular dynamics (AIMD) simulations using an NPT ensemble.

## 5. Summary

In summary, by combining CGCNN ML approach with first-principles DFT calculations, we predict five new Ce-Co-Cu ternary compounds ($Ce_3Co_3Cu$, $CeCoCu_2$, $Ce_{12}Co_7Cu$, $Ce_{11}Co_9Cu$ and $Ce_{10}Co_{11}Cu_4$) that are both thermodynamically and dynamically stable. These compounds are Ce rich, and thus likely not good for permanent magnet applications. However, these compounds might be interesting if the Ce is mixed valent, a question that cannot be examined at the level of DFT calculations, instead requiring further investigation with methods like DMFT. Moreover, our ML-guided approach identifies two Co-rich low-energy compounds ($Ce_4Co_{33}Cu$ and $Ce_4Co_{31}Cu_3$) that exhibit high magnetizations, and where $Ce_4Co_{31}Cu_3$ has Cu occuping a Co-dumbell that likely leads to enhanced magnetic anisotropy. The composition region circled by the red line that encloses these two compounds in Fig. 4 might be a promising place to look experimentally for compounds suitable for permanent magnet applications.

The search for stable and low-energy ternary compounds presents a vast and complex space due to the numerous possible combinations of compositions and crystal structures. The CGCNN ML approach in our ML-guided framework efficiently navigates through this vast space. In just four hours on a standard GPU, CGCNN evaluated the energies and magnetic moments of 854,070 hypothetical Ce-Co-Cu compounds, leading to the identification of 5,719 (~0.7% of the total) viable structures spanning over 2,522 compositions for further DFT analysis. This approach significantly accelerates the discovery process. However, it is important to note that the CGCNN+DFT approach does not guarantee exhaustive coverage, as some stable and low-energy low-energy phases may be overlooked if their structural templates are absent from the CGCNN database. In addition, future studies should also consider disordered structures with partial occupancy. Nonetheless, this ML-guided framework represents a powerful tool for rapidly identifying stable compounds across vast chemical and structural spaces, marking a new paradigm in materials design and discovery in the digital era, poised for broad adoption and impact.

**Data availability**

The data leading to the findings in this paper and the ML models are available from the authors upon reasonable request.

**Acknowledgments**

Work at Ames National Laboratory was supported by the U.S. Department of Energy (DOE), Office of Science, Basic Energy Sciences, Materials Science and Engineering Division including a grant of computer time at the National Energy Research Supercomputing Center (NERSC) in Berkeley. Ames National Laboratory is operated for the U.S. DOE by Iowa State University under contract # DE-AC02-07CH11358.

# Supplementary material

**Supplementary Table 1.** Structure database of the Ce-Co-Cu system, including existing phases, ML+DFT predicted stable phases and low-energy phases. The formation energies ($E_f$) are referenced to the stable phases that form Gibbs triangles.

| | Phases | Symmetry | a (Å) | b (Å) | c (Å) | $E_f$ (meV/atom) |
|---|---|---|---|---|---|---|
| **Existing stable phases** | **Ce** | *C2/m* | 5.77 | 5.77 | 3.32 | **0** |
| | **Co** | *P*6$_3$*/mmc* | 2.49 | 2.49 | 4.02 | **0** |
| | **Cu** | *Fm*-3*m* | 3.62 | 3.62 | 3.62 | **0** |
| | **$CeCo_2$** | *Fd-3m* | 7.07 | 7.07 | 7.07 | **0** |
| | **$Ce_2Co_{17}$** | *R-3m* | 8.36 | 8.36 | 12.16 | **0** |
| | **$CeCu_2$** | *Imma* | 4.40 | 6.64 | 7.35 | **0** |
| | **$CeCu_6$** | *P*2$_1$*/c* | 5.03 | 8.06 | 10.10 | **0** |
| **ML+DFT predicted stable phase** | **$Ce_3Co_3Cu$** | *Cmcm* | 5.33 | 5.33 | 5.33 | **0** |
| | **$CeCoCu_2$** | *Pnma* | 7.44 | 6.95 | 4.96 | **0** |
| | **$Ce_{12}Co_7Cu$** | *I4/mcm* | 11.50 | 11.50 | 12.81 | **0** |
| | **$Ce_{11}Co_9Cu$** | *Iba2* | 11.01 | 11.74 | 12.73 | **0** |
| | **$Ce_{10}Co_{11}Cu_4$** | *C2/m* | 10.07 | 4.32 | 20.25 | **0** |
| Existing low-energy phases | $Ce_2Co_{16}Cu$ | *P-3m1* | 8.36 | 8.36 | 8.12 | 5 |
| | $CeCo_4Cu$ | *P-6m2* | 4.93 | 4.93 | 4.04 | 24 |
| | $CeCo_3Cu_2$ | *P6/mmm* | 4.94 | 4.94 | 4.02 | 38 |
| | $Ce_2Co_5Cu_5$ | *Pmmm* | 4.10 | 4.88 | 8.46 | 44 |
| ML+DFT predicted low-energy phases | CeCoCu | *P4/nmm* | 3.62 | 3.62 | 7.67 | 8 |
| | Ce4Co3Cu31 | *P1* | 6.40 | 6.45 | 12.86 | 16 |
| | Ce2CoCu4 | *P4/nmm* | 6.89 | 6.89 | 4.88 | 18 |
| | CeCoCu2 | *Pnma* | 7.23 | 6.96 | 4.98 | 21 |
| | Ce3(CoCu2)2 | *P-62m* | 6.81 | 6.81 | 3.85 | 22 |
| | Ce4CoCu2 | *P2_1/c* | 9.18 | 6.39 | 20.34 | 23 |
| | Ce4Co31Cu3 | *P1* | 6.29 | 6.30 | 12.59 | 25 |
| | Ce3Co2Cu | *P-4m2* | 4.18 | 4.18 | 6.46 | 26 |
| | Ce3(CoCu)2 | *Pnma* | 15.55 | 4.64 | 7.16 | 26 |
| | Ce16Co10Cu | *P-4* | 8.41 | 8.41 | 8.26 | 27 |
| | Ce2CoCu16 | *P-3m1* | 8.60 | 8.60 | 8.16 | 27 |
| | Ce4CoCu3 | *Pmm2* | 3.31 | 4.49 | 10.15 | 27 |
| | CeCoCu | *P2_1/c* | 5.90 | 4.92 | 7.57 | 28 |
| | Ce4Co3Cu5 | *P-6m2* | 5.17 | 5.17 | 8.29 | 30 |
| | Ce4CoCu7 | *P-3m1* | 5.13 | 5.13 | 8.72 | 30 |
| | Ce4CoCu33 | *P3m1* | 8.50 | 8.50 | 8.21 | 30 |
| | Ce2CoCu3 | *P6_3/mmc* | 5.01 | 5.01 | 9.03 | 31 |
| | Ce2CoCu16 | *P-3m1* | 8.52 | 8.52 | 8.14 | 33 |
| | CeCoCu | *Pnma* | 6.59 | 4.52 | 6.91 | 34 |
| | Ce2CoCu3 | *Pmc2_1* | 4.39 | 7.28 | 6.64 | 37 |
| | Ce6CoCu11 | *P6_3/mcm* | 8.97 | 8.97 | 8.69 | 37 |
| | Ce4CoCu | *Pmc2_1* | 4.44 | 9.36 | 6.30 | 38 |
| | Ce2CoCu | *P6_3cm* | 7.60 | 7.60 | 9.04 | 38 |
| | Ce4Co2Cu3 | *P2/m* | 6.94 | 3.53 | 7.47 | 38 |
| | Ce2CoCu3 | *P6_3/mmc* | 5.02 | 5.02 | 8.76 | 40 |
| | CeCoCu2 | *Pnma* | 6.84 | 7.23 | 5.27 | 40 |
| | CeCoCu | *Pnma* | 6.61 | 3.70 | 8.13 | 41 |
| | Ce2Co2Cu | *P2/m* | 3.66 | 3.65 | 7.08 | 41 |
| | Ce3CoCu3 | *P2_1/m* | 4.55 | 9.37 | 6.26 | 42 |
| | Ce4Co5Cu29 | *P-1* | 6.42 | 6.45 | 12.86 | 43 |
| | CeCoCu4 | *P-6m2* | 4.95 | 4.95 | 4.13 | 43 |
| | Ce6CoCu | *Pmc2_1* | 9.59 | 6.16 | 6.73 | 44 |
| | Ce3Co2Cu3 | *Pnma* | 7.63 | 15.34 | 5.05 | 44 |
| | Ce5Co2Cu | *P2_1/c* | 8.07 | 8.02 | 11.06 | 44 |
| | Ce5(CoCu2)2 | *Pbam* | 7.25 | 16.93 | 3.47 | 45 |
| | Ce3Co2Cu | *P2_1/c* | 8.69 | 6.81 | 15.69 | 45 |
| | Ce3Co2Cu3 | *Pnma* | 7.62 | 15.22 | 5.06 | 45 |
| | Ce4Co7Cu | *P-3m1* | 5.07 | 5.07 | 8.10 | 47 |
| | Ce6CoCu2 | *P-62m* | 8.28 | 8.28 | 3.38 | 47 |
| | Ce7Co2Cu | *P4/mbm* | 11.63 | 11.63 | 3.54 | 48 |
| | Ce4Co29Cu5 | *P-1* | 6.30 | 6.31 | 12.62 | 48 |
| | CeCoCu | *Pnma* | 13.24 | 3.74 | 8.04 | 48 |
| | Ce5Co2Cu | *P2_1/c* | 8.13 | 7.76 | 11.33 | 48 |
| ML+DFT predicted low-energy phases | Ce4Co5Cu3 | *P-6m2* | 5.01 | 5.01 | 8.60 | 48 |
| (continued) | Ce4Co7Cu | *P-3m1* | 5.06 | 5.06 | 8.11 | 48 |
| | Ce20Co12Cu | *P2/m* | 6.68 | 13.76 | 7.62 | 48 |
| | Ce4CoCu3 | *P-1* | 5.57 | 6.11 | 10.52 | 49 |
| | Ce4Co5Cu3 | *P-6m2* | 5.02 | 5.02 | 8.52 | 49 |
| | Ce2CoCu3 | *P6_3/mmc* | 5.01 | 5.01 | 8.67 | 49 |
| | Ce3(CoCu)2 | *P2_1/m* | 5.05 | 13.25 | 12.00 | 49 |
| | Ce2Co2Cu | *P4/mbm* | 6.98 | 6.98 | 3.63 | 49 |
| | CeCo2Cu3 | *P6/mmm* | 8.61 | 8.61 | 4.03 | 50 |
| | CeCoCu | *Pnma* | 6.50 | 3.89 | 7.84 | 50 |
| | Ce24Co14Cu3 | *P1* | 7.19 | 7.19 | 18.88 | 50 |
| | CeCoCu | *P2_1/m* | 6.85 | 3.70 | 11.88 | 51 |
| | Ce5(CoCu)2 | *Pbam* | 6.48 | 12.81 | 4.44 | 51 |
| | Ce3(Co2Cu)2 | *P-1* | 4.94 | 5.04 | 12.64 | 51 |
| | Ce10Co6Cu | *P1* | 7.19 | 7.20 | 30.69 | 51 |
| | Ce3Co4Cu | *Pm* | 4.83 | 4.28 | 20.40 | 52 |
| | Ce3Co2Cu | *P2_1/c* | 8.81 | 6.80 | 15.51 | 52 |
| | Ce5Co2Cu | *P4/ncc* | 11.13 | 11.13 | 5.70 | 52 |
| | Ce3(CoCu)2 | *P2_1/m* | 5.02 | 13.21 | 12.01 | 53 |
| | Ce6Co11Cu | *P6_3/mcm* | 8.75 | 8.75 | 8.14 | 53 |
| | Ce20(Co3Cu)3 | *P422* | 7.54 | 7.54 | 12.01 | 53 |
| | Ce3Co3Cu4 | *Pnma* | 23.42 | 4.01 | 7.09 | 53 |
| | Ce6Co4Cu7 | *Pbcm* | 5.14 | 7.63 | 31.33 | 53 |
| | Ce5(CoCu2)2 | *Pbam* | 7.31 | 16.94 | 3.34 | 54 |
| | Ce6Co11Cu | *P6_3/mcm* | 8.75 | 8.75 | 8.11 | 54 |
| | Ce2CoCu11 | *Pmc2_1* | 5.05 | 9.90 | 8.09 | 54 |
| | CeCo4Cu | *P6/mmm* | 8.51 | 8.51 | 3.99 | 54 |
| | Ce2Co2Cu | *P4/mbm* | 7.04 | 7.04 | 3.54 | 54 |
| | Ce4Co3Cu | *Pm* | 5.41 | 4.35 | 6.76 | 54 |
| | CeCoCu | *Pnma* | 7.00 | 3.78 | 15.41 | 55 |
| | Ce4CoCu2 | *P-1* | 4.13 | 5.37 | 7.31 | 55 |
| | Ce3Co2Cu | *Pm* | 4.57 | 3.96 | 6.38 | 55 |
| | CeCoCu | *Pbcm* | 5.26 | 7.82 | 4.97 | 55 |
| | Ce4Co5Cu3 | *P-6m2* | 5.01 | 5.01 | 8.39 | 56 |
| | Ce18Co5Cu4 | *Pm* | 8.19 | 3.48 | 21.61 | 56 |
| | CeCoCu4 | *P-6m2* | 4.95 | 4.95 | 4.02 | 56 |
| | Ce4CoCu3 | *Pnma* | 7.08 | 8.90 | 10.76 | 56 |
| | Ce3Co3Cu | *Pnma* | 12.77 | 3.50 | 11.33 | 56 |
| | Ce2Co2Cu15 | *P6_3/mmc* | 8.58 | 8.58 | 8.14 | 56 |
| | Ce3Co3Cu4 | *Pnma* | 16.55 | 3.74 | 10.36 | 57 |
| | Ce4Co3Cu5 | *P-6m2* | 5.08 | 5.08 | 8.14 | 57 |
| | Ce3(CoCu)2 | *Pbcm* | 5.42 | 8.04 | 12.16 | 57 |
| | Ce4CoCu | *P4/mcc* | 6.94 | 6.94 | 5.55 | 57 |
| | Ce6Co4Cu | *P6_3mc* | 9.00 | 9.00 | 6.45 | 58 |
| | Ce10Co6Cu | *P1* | 7.25 | 7.29 | 30.15 | 58 |
| | Ce4Co4Cu | *Pnma* | 7.07 | 13.85 | 6.92 | 59 |
| | CeCoCu4 | *P6/mmm* | 4.99 | 4.99 | 8.21 | 59 |
| | Ce3(CoCu)2 | *P2_1/m* | 5.23 | 12.68 | 12.15 | 59 |
| | Ce2CoCu14 | *P4_2/mnm* | 8.74 | 8.74 | 12.57 | 59 |
| | CeCoCu4 | *P6/mmm* | 8.71 | 8.71 | 3.99 | 59 |
| | CeCoCu | *P2_1/m* | 6.69 | 3.78 | 11.58 | 60 |
| ML+DFT predicted low-energy phases (continued) | Ce4Co3Cu | *Pmm2* | 3.74 | 4.30 | 9.93 | 60 |
| | CeCoCu | *P2_1/c* | 5.96 | 5.77 | 8.11 | 61 |
| | Ce3Co3Cu | *Pnma* | 16.24 | 3.61 | 8.65 | 61 |
| | CeCo2Cu | *Pnma* | 6.74 | 4.20 | 9.03 | 62 |
| | Ce4Co4Cu | *Pnma* | 7.00 | 13.79 | 6.92 | 62 |
| | Ce2CoCu | *Pnma* | 7.89 | 8.10 | 4.55 | 63 |
| | Ce3CoCu | *C2/m* | 6.61 | 10.33 | 6.79 | 63 |
| | Ce3Co3Cu | *P2_1/m* | 4.86 | 9.63 | 5.67 | 63 |
| | Ce3Co2Cu | *P2_1/m* | 5.51 | 4.36 | 10.14 | 64 |
| | Ce3Co2Cu | *P2_1/m* | 5.51 | 4.36 | 10.14 | 64 |
| | Ce4Co3Cu | *Pmm2* | 3.61 | 4.43 | 9.80 | 64 |
| | Ce2Co3Cu | *P6_3/mmc* | 5.15 | 5.15 | 8.01 | 64 |
| | Ce2Co3Cu | *P6_3/mmc* | 5.14 | 5.14 | 8.00 | 64 |
| | Ce4Co3Cu | *Pmm2* | 3.73 | 4.22 | 10.11 | 65 |
| | Ce3CoCu2 | *P2_1/m* | 5.79 | 13.29 | 9.82 | 65 |
| | Ce5(CoCu)2 | *P2_1* | 7.55 | 23.95 | 8.53 | 65 |
| | Ce2Co3Cu | *P6_3/mmc* | 5.12 | 5.12 | 8.01 | 65 |
| | CeCo2Cu | *P6_3/mmc* | 4.30 | 4.30 | 8.23 | 65 |
| | Ce4Co3Cu | *Pmm2* | 3.32 | 4.26 | 10.41 | 66 |
| | CeCo2Cu | *Pmma* | 4.72 | 4.08 | 6.41 | 66 |
| | CeCoCu | *P2_1/c* | 5.85 | 5.27 | 7.13 | 66 |
| | Ce5(CoCu3)2 | *Pm-3* | 9.06 | 9.06 | 9.06 | 66 |
| | CeCoCu | *Pmmn* | 3.77 | 27.11 | 6.91 | 66 |
| | Ce3(CoCu)2 | *Pbcm* | 5.12 | 8.03 | 12.89 | 66 |
| | Ce2Co3Cu | *P6_3/mmc* | 5.11 | 5.11 | 7.99 | 67 |
| | Ce3CoCu3 | *P2_1/c* | 5.85 | 15.31 | 8.19 | 67 |
| | Ce4Co3Cu | *P-31m* | 7.72 | 7.72 | 6.01 | 67 |
| ML+DFT predicted low-energy phases | CeCoCu2 | *Pnma* | 13.91 | 7.30 | 5.04 | 67 |
| (continued) | | | | | | |
| | Ce10Co2Cu19 | *P6/mmm* | 14.49 | 14.49 | 8.88 | 68 |
| | Ce2Co3Cu14 | *P6_3/mmc* | 8.53 | 8.53 | 8.23 | 68 |
| | CeCoCu | *Pmmn* | 3.82 | 19.63 | 6.74 | 68 |
| | Ce10Co5Cu3 | *P2_1/c* | 8.65 | 7.80 | 23.59 | 68 |
| | Ce5Co3Cu | *P2_1/c* | 15.83 | 8.73 | 10.69 | 69 |
| | CeCo2Cu3 | *P6/mmm* | 8.69 | 8.69 | 4.03 | 69 |
| | CeCo2Cu | *Pmma* | 4.72 | 4.06 | 6.42 | 69 |
| | Ce2Co3Cu14 | *P6_3/mmc* | 8.50 | 8.50 | 8.19 | 69 |
| | Ce20(Co4Cu)3 | *Pmm2* | 4.27 | 25.29 | 6.70 | 69 |
| | Ce3(CoCu)2 | *Pbcm* | 5.52 | 7.06 | 13.44 | 70 |
| | Ce14Co16Cu | *Pmn2_1* | 16.49 | 6.17 | 10.98 | 70 |
| | Ce2(CoCu)5 | *Pm* | 8.69 | 4.03 | 14.87 | 70 |
| | Ce4CoCu7 | *P-3m1* | 5.09 | 5.09 | 8.33 | 71 |
| | Ce3(CoCu2)2 | *Pnma* | 12.36 | 4.32 | 11.53 | 71 |
| | Ce4CoCu3 | *Pm* | 5.53 | 4.46 | 6.92 | 71 |
| | Ce5Co3Cu | *P2_1/c* | 8.29 | 6.66 | 15.24 | 71 |
| | Ce5Co4Cu | *P2/m* | 7.23 | 7.23 | 7.33 | 72 |
| | Ce2Co3Cu7 | *P2/m* | 5.00 | 4.03 | 8.72 | 72 |
| | Ce10Co7Cu3 | *Pmc2_1* | 3.74 | 9.49 | 22.87 | 72 |
| | Ce2Co3Cu | *P6_3/mmc* | 5.05 | 5.05 | 8.07 | 73 |
| | Ce3CoCu | *P2_1/c* | 6.45 | 9.71 | 7.05 | 73 |
| | Ce3(CoCu)2 | *Pbcm* | 5.47 | 7.14 | 13.20 | 73 |
| | Ce4CoCu3 | *Pm* | 5.49 | 4.43 | 6.95 | 74 |
| | Ce7Co4Cu | *P2_1* | 9.21 | 6.82 | 9.32 | 74 |
| | Ce5Co2Cu | *P2_1/c* | 8.94 | 6.50 | 12.18 | 74 |
| | Ce3CoCu5 | *P-62m* | 6.67 | 6.67 | 4.17 | 75 |
| | Ce15Co9Cu | *P6_3mc* | 13.60 | 13.60 | 6.58 | 75 |
| | Ce15(Co2Cu5)4 | *Pm* | 7.10 | 7.12 | 31.03 | 75 |
| | CeCoCu | *P2_1/c* | 6.07 | 6.00 | 6.48 | 75 |
| | CeCo2Cu3 | *P-6m2* | 8.65 | 8.65 | 4.06 | 76 |
| | Ce3CoCu5 | *Pnma* | 11.16 | 8.29 | 7.09 | 76 |
| | CeCoCu | *P-62m* | 6.83 | 6.83 | 3.78 | 78 |
| | Ce2CoCu4 | *Pnma* | 9.04 | 3.91 | 13.25 | 78 |
| | Ce3(CoCu2)2 | *P-1* | 5.17 | 5.28 | 12.16 | 78 |
| | CeCoCu | *P-62c* | 6.83 | 6.83 | 7.57 | 79 |
| | CeCoCu | *P-62m* | 6.78 | 6.78 | 3.82 | 79 |
| | Ce5CoCu3 | *P2_1/c* | 6.78 | 14.82 | 8.72 | 79 |
| | Ce3CoCu2 | *P2_1/m* | 5.50 | 4.47 | 10.40 | 79 |
| | Ce2Co3Cu | *P-6m2* | 5.20 | 5.20 | 4.32 | 79 |
| | Ce8CoCu5 | *Pc* | 7.14 | 18.91 | 11.57 | 79 |
| | CeCoCu | *P-62m* | 6.79 | 6.79 | 3.76 | 79 |
| | Ce5Co4Cu | *P-4n2* | 7.33 | 7.33 | 7.09 | 79 |
| | Ce5Co4Cu | *P4/m* | 7.13 | 7.13 | 3.72 | 80 |
| | Ce2CoCu | *P2_1/m* | 5.62 | 4.00 | 7.65 | 80 |
| | Ce2CoCu3 | *P-1* | 5.60 | 5.60 | 8.00 | 80 |
| | Ce3(CoCu)2 | *Pbcm* | 5.33 | 7.47 | 13.10 | 80 |
| | Ce2CoCu3 | *Pm* | 4.32 | 3.55 | 7.33 | 80 |
| | Ce4Co3Cu2 | *P-1* | 5.20 | 7.50 | 9.48 | 81 |
| ML+DFT predicted low-energy phases (continued) | CeCoCu | *Pbcm* | 6.14 | 4.72 | 7.48 | 81 |
| | Ce2CoCu2 | *Pnma* | 6.44 | 8.74 | 6.79 | 82 |
| | Ce3Co2Cu3 | *P6_3/mmc* | 5.16 | 5.16 | 12.17 | 82 |
| | CeCoCu | *P-62c* | 6.87 | 6.87 | 7.37 | 82 |
| | Ce2CoCu3 | *Pmm2* | 3.55 | 4.33 | 7.26 | 82 |
| | Ce3CoCu2 | *P-3m1* | 4.82 | 4.82 | 5.98 | 82 |
| | Ce4CoCu2 | *P-1* | 5.18 | 6.30 | 9.39 | 83 |
| | Ce2(CoCu)5 | *Pmmm* | 4.04 | 4.93 | 8.71 | 83 |
| | Ce2CoCu3 | *P-1* | 3.74 | 7.86 | 8.04 | 83 |
| | Ce3(CoCu2)2 | *P-1* | 5.07 | 5.15 | 12.42 | 83 |
| | Ce3Co2Cu3 | *Pnma* | 7.96 | 14.18 | 4.91 | 84 |
| | CeCoCu | *Cmcm* | 4.34 | 9.04 | 4.99 | 84 |
| | Ce4CoCu3 | *Pm* | 3.82 | 4.30 | 10.49 | 84 |
| | Ce2CoCu | *Pmc2_1* | 4.49 | 5.41 | 6.82 | 84 |
| | CeCoCu | *P-62m* | 6.68 | 6.68 | 3.95 | 85 |
| | Ce2CoCu3 | *P6_3/mmc* | 8.77 | 8.77 | 6.36 | 85 |
| | Ce5Co3Cu | *Pnma* | 18.56 | 3.73 | 10.90 | 85 |
| | Ce5CoCu4 | *P4/mmm* | 3.32 | 3.32 | 18.24 | 85 |
| | Ce2CoCu3 | *P6_3/mmc* | 4.21 | 4.21 | 14.37 | 85 |
| | Ce2CoCu3 | *P-6m2* | 4.25 | 4.25 | 7.12 | 86 |
| | Ce4CoCu3 | *Pmm2* | 3.78 | 4.29 | 10.61 | 86 |
| | Ce5Co2Cu | *Pbam* | 8.74 | 11.13 | 7.54 | 87 |
| | Ce6Co3Cu | *P1* | 7.30 | 7.95 | 8.44 | 87 |
| | Ce3Co3Cu8 | *P2/m* | 8.74 | 4.02 | 12.49 | 88 |
| | Ce2CoCu | *Pmc2_1* | 4.16 | 5.22 | 7.53 | 88 |
| | Ce2CoCu3 | *Pmm2* | 3.53 | 4.19 | 7.43 | 88 |
| | Ce4CoCu | *Pmn2_1* | 3.56 | 11.78 | 13.08 | 88 |
| | Ce3(CoCu2)2 | *P-6* | 7.11 | 7.11 | 3.59 | 89 |
| | Ce3CoCu2 | *P4/mmm* | 3.38 | 3.38 | 10.75 | 89 |
| | CeCoCu2 | *Pmma* | 4.78 | 3.95 | 6.53 | 90 |
| | Ce4CoCu3 | *P4/mmm* | 3.51 | 3.51 | 13.34 | 90 |
| | Ce4CoCu3 | *Pmm2* | 3.98 | 4.21 | 10.33 | 90 |
| | Ce2(CoCu)5 | *Pmmm* | 4.11 | 4.86 | 8.54 | 90 |
| | Ce2CoCu3 | *P6_3/mmc* | 8.64 | 8.64 | 6.39 | 90 |
| | Ce3Co4Cu | *Pm* | 4.94 | 4.12 | 6.86 | 91 |
| | CeCoCu | *Pnma* | 6.92 | 4.64 | 6.45 | 91 |
| | Ce2CoCu | *P-1* | 6.78 | 7.28 | 7.29 | 91 |
| | CeCoCu | *P2_1/m* | 6.72 | 4.58 | 6.79 | 91 |
| | Ce2CoCu3 | *Pmm2* | 3.13 | 4.42 | 7.59 | 91 |
| | CeCoCu | *P2_1/c* | 5.69 | 6.34 | 7.02 | 91 |
| | Ce7Co2Cu3 | *P2_1/c* | 10.02 | 6.12 | 17.83 | 91 |
| | CeCoCu4 | *Pmma* | 7.10 | 3.94 | 6.41 | 91 |
| | Ce3(CoCu2)2 | *P3m1* | 4.24 | 4.24 | 21.34 | 92 |
| | Ce4Co4Cu | *P2_1/c* | 7.40 | 7.27 | 13.34 | 92 |
| | Ce3(CoCu2)2 | *P-6* | 7.14 | 7.14 | 3.55 | 92 |
| | Ce3CoCu2 | *Pnma* | 11.68 | 4.39 | 9.78 | 92 |
| | Ce2(CoCu)5 | *Pmmm* | 4.08 | 4.86 | 8.48 | 92 |
| | Ce4CoCu | *P1* | 7.10 | 7.13 | 8.65 | 93 |
| | Ce2CoCu3 | *Pmm2* | 3.12 | 4.42 | 7.63 | 93 |
| ML+DFT predicted low-energy phases (continued) | Ce3CoCu | *P2_1/c* | 6.89 | 7.09 | 10.52 | 93 |
| | Ce2Co5Cu12 | *P6_3/mmc* | 8.52 | 8.52 | 8.15 | 93 |
| | Ce7Co8Cu3 | *P2_1/c* | 14.15 | 6.74 | 15.32 | 93 |
| | Ce3(CoCu2)2 | *P3m1* | 4.22 | 4.22 | 21.12 | 93 |
| | Ce4CoCu3 | *Pmm2* | 3.66 | 4.27 | 10.51 | 94 |
| | Ce2CoCu | *Pmc2_1* | 4.04 | 4.36 | 9.47 | 94 |

| | | | | | |
|---|---|---|---|---|---|
| Ce3Co2Cu3 | *Pm* | 3.73 | 3.69 | 10.92 | 94 |
| Ce2CoCu | *Pmc2_1* | 4.01 | 5.23 | 7.82 | 94 |
| Ce2Co5Cu12 | *P6_3/mmc* | 8.49 | 8.49 | 8.14 | 94 |
| Ce3(CoCu2)2 | *Pnma* | 12.53 | 4.36 | 11.33 | 95 |
| Ce7Co3Cu13 | *Pnma* | 11.58 | 20.41 | 6.66 | 95 |
| Ce4Co2Cu | *P2_1/c* | 6.74 | 7.10 | 13.05 | 95 |
| Ce3(CoCu2)2 | *P-1* | 5.07 | 5.12 | 12.39 | 95 |
| Ce4CoCu | *P2_1/c* | 7.00 | 8.33 | 11.17 | 95 |
| Ce4Co2Cu | *P-1* | 4.15 | 5.46 | 6.64 | 95 |
| CeCo3Cu2 | *P-62m* | 8.63 | 8.63 | 3.97 | 95 |
| Ce10Co3Cu7 | *Pmc2_1* | 3.73 | 9.55 | 22.77 | 96 |
| Ce5CoCu3 | *P2_1/c* | 10.19 | 12.87 | 5.79 | 96 |
| CeCo2Cu3 | *P6/mmm* | 4.85 | 4.85 | 4.18 | 96 |
| Ce9Co3Cu2 | *P-1* | 6.43 | 6.66 | 15.06 | 96 |
| Ce2CoCu3 | *P6_3/mmc* | 8.40 | 8.40 | 7.34 | 97 |
| Ce3CoCu | *Pnma* | 5.20 | 6.91 | 12.28 | 97 |
| Ce2CoCu | *P1* | 8.60 | 8.83 | 9.00 | 97 |
| Ce8Co3Cu | *P2* | 5.55 | 28.78 | 5.55 | 97 |
| CeCoCu | *Pnma* | 6.78 | 4.55 | 6.56 | 98 |
| Ce2CoCu3 | *P-62m* | 9.25 | 9.25 | 4.19 | 98 |
| Ce3Co2Cu | *Pnma* | 11.12 | 4.11 | 10.46 | 98 |
| Ce3CoCu | *P6_3/mmc* | 6.97 | 6.97 | 5.24 | 99 |
| Ce5(CoCu)4 | *P-1* | 3.92 | 7.86 | 7.88 | 99 |
| Ce2CoCu | *P4/mmm* | 3.46 | 3.46 | 6.76 | 99 |
| Ce3CoCu | *Pnma* | 7.94 | 9.93 | 5.48 | 99 |